\documentclass[12pt, a4paper]{article}
\usepackage[utf8]{inputenc}

\usepackage{float}
\usepackage{amsmath}
\usepackage{amssymb}
\usepackage{multirow}
\usepackage{booktabs}
\usepackage[T1]{fontenc}
\usepackage{graphicx}
\usepackage{xcolor}

\DeclareMathOperator*{\argmin}{arg\,min}

\begin{document}

\title{A Bayesian analysis of gain-loss asymmetry}

\date{\today}

\author{Andrea Di Iura$^{\dagger}$, Giulia Terenzi$^{\star}$ \\
$^{\dagger}$\texttt{andreadiiura@gmail.com}, $^{\star}$\texttt{terenzi.giu@gmail.com}}
\vspace{10pt}
\maketitle

\begin{abstract}
We perform a quantitative analysis of the gain/loss asymmetry for financial time series by using a Bayesian approach. In particular, we focus on some selected indices and analyze the statistical significance of the asymmetry amount through a Bayesian generalization of the t-Test, which relaxes the normality assumption on the underlying distribution. We propose two different models for data distribution, we study the convergence of our method and we provide several graphical representations of our numerical results. Finally, we perform a sensitivity analysis with respect to model parameters in order to study the reliability and robustness of our results.
\end{abstract}

\section{Introduction}

Financial time-series have been widely studied in the scientific literature. 
It is well established that this kind of time-series have common features shared among different asset classes that can not be achieved by classical models such as the Black and Scholes Geometrical Brownian Motion \cite{black1973pricing}. The same F. Black, in 1989, wrote about the holes in the worldwide used model \cite{black1989holes} which bears his name. Those properties are the so-called stylized facts \cite{cont2001empirical} (see also Refs. \cite{chakraborti2011econophysics, jiang2019multifractal} for a review on this subject). Among these characteristics, for example, we recall the well-known fact that the autocovariance function of the absolute returns shown a power-law decay \cite{cont1997scaling, gopikrishnan1999scaling, liu1997correlations} or that several instruments have a weak correlation of the returns \cite{malkiel1970efficient, pagan1996econometrics, cont1997scaling}.

One of the well-known stylized facts regarding financial time series is the so-called \emph{gain/loss asymmetry} \cite{zhang1999toward, jensen2003inverse}. Roughly speaking, it consists of the fact that the drawdowns observed in stock prices and stock index values are not as large as the upward movements. This effect was widely studied in the literature to analyze stocks \cite{zaluska2006comparison}, indices \cite{siven2009multiscale, sandor2016time} as well as stocks in emerging markets \cite{karpio2007gain}. Many attempts to deal with this effect have been considered, such as an analysis of the clustering on the asymmetry properties in financial time series \cite{tseng2011asset}, symmetry breaking mechanisms \cite{savona2015financial}, or by using empirical and agents modeled studies \cite{yamashita2017detection, peiro2004asymmetries}. Moreover, this property is particularly interesting since it can be used by practitioners to achieve a desired level of profit using the optimal investment horizon \cite{jensen2003inverse} or to hedge the risk in a proper way \cite{pruna2016asset}. According to this empirical evidence, many models have been proposed. As an example, in \cite{donangelo2006synchronization} a collective mechanism has been considered, where the fear of practitioners is stronger than their optimism on average. This is consistent with economic models in which the utility loss of negative returns is larger than the utility gain for positive returns in the case of most investors \cite{kahneman2013prospect}. 

This paper aims to quantitatively investigate the gain/loss asymmetry by using a Bayesian approach for few selected indices. In particular, we analyze how statistically significant is the amount of asymmetry through a Bayesian generalization of the t-Test, \cite{kruschke2013bayesian} developed to compare two populations without any normality assumptions on the underlying distribution. Using this tool we are able to study the gain/loss effect and to quantify the impact of the model parameter on the robustness of the results.

In our analysis, we focus on indices, since it is known in the literature that the gain/loss asymmetry is not so evident in the stocks market \cite{balogh2010persistent}. In particular, we have selected ten indices from a few countries and markets which are generally recognized as good economic benchmarks.
We propose two different models for data distribution: the standard Student-$t$ distribution and an Inverse Gamma distribution, inspired by the distribution of the first hitting time of the Geometrical Brownian Motion.  For both models, we study the convergence of the Markov Chain Monte Carlo (MCMC) method used in the analysis and we present our numerical results discussing them with the support of several graphical representations. Moreover, in order to assert the reliability of our results we perform a sensitivity analysis with respect to the model parameters and the time interval considered for the underlying time series.

We get quantitative evidence that confirms the gain/loss asymmetry for most of the financial series taken into account. 
Therefore, our results provide a quantitative analysis of the  gain/loss asymmetry property which turns out to be robust and reliable. As far as we know, this kind of accurate and solid investigation was missing in the literature. 

The paper is organized as follows: in Section \ref{sec:methodology} we set our framework and discuss the adopted methodology and the used Bayesian tools. In Section \ref{sec:dataset} are analyzed the main features of the considered dataset, while in Section \ref{sec:section-results} we present the numerical results of the performed analysis and the sensitivity study. We conclude in Section \ref{sec:conclusion}, summarising the most important results and providing additional remarks.

\section{Bayesian analysis}
\label{sec:methodology}
Let $S_t$ denote the (detrended) asset price.
 It is well-known that financial time series are usually non-stationary and often characterized by the presence of a positive drift over long time scales.  The presence of such a drift does not allow to directly compare the distribution of positive and negative levels of return. In \cite{simonsen2002optimal} the authors use a wavelet transformation to decompose the time series in a trend and signal components, while in \cite{karpio2007gain} a moving average of 100 days have been used. In our approach, we subtract to the price series the moving median over a certain number of days, which we call \textit{filter size} $f$. This quantity will be the object of sensitivity analysis in Section \ref{sec:section-results}. With respect to the moving mean, the median is less sensitive to outliers, thus we are able to capture tiny movements in the detrended time-series.

 As usual, returns are defined as the difference of logarithms of successive prices, i.e. 
\begin{equation}
\label{eq:returns_definition}
r_t = \ln{S_t}-    \ln{S_{t-\Delta t}}. 
\end{equation}
In our case, the time scale $\Delta t$ is equal to one business day.

We use the formulation proposed in Refs. \cite{simonsen2002optimal, jensen2003inverse}, where the authors define an observable random variable $\tau_{\pm}$ that quantify the amount of asymmetry. For a given return level $\rho >0$, it is defined as the following stopping time
\begin{equation}
\label{eq:tau_definition}
\begin{cases}
\tau_+ = \argmin_{\tau>0} ( r_\tau \geq \rho ),\\ \tau_- = \argmin_{\tau>0} ( r_\tau \leq -\rho ).
\end{cases}    
\end{equation}
From a purely probabilistic point of view, this is the First Hitting  Time of the process $r_t$ for a given barrier $\rho$.
This process has been widely studied in the literature for different stochastic processes describing the dynamics of the underlying process $S_{t\geq 0}$ (see, for example, \cite{alili2005representations, yi2010first, valenti2007hitting, masoliver2009first}). In particular, for the Geometrical Brownian Motion it can be easily proved that it exists a closed form solution for the distribution of $\tau_{\pm }$. In this framework, the diffusion process $S_{t\geq 0}$  is driven by the stochastic differential equation ${\rm d}S_t = S_t(\lambda{\rm d}t + \Sigma {\rm d}B_t)$, where $\lambda$ is the drift coefficient, $\Sigma$ the diffusion coefficient and $B_t$ is a Wiener process.
Then, the log-price process reduces to a Brownian Motion with drift, and the first hitting stopping time is distributed  as an Inverse Gamma Distribution that depends on the parameter of the stochastic differential equation, of the initial value $r_0=\log S_0$ and of the crossing level $\rho >0$, with probability density function
\begin{equation}
p(t) = \sqrt{\frac{(\rho-r_0)^2}{2\pi\Sigma^2t^3}} \exp \left[ - \frac{(\lambda t - \rho +r_0)^2 }{2\Sigma^2 t} \right].
\end{equation}
This is a generalization of the well-known Reflection Principle for Brownian Motion. For a complete proof, we refer,  for example, to \cite{schilling2012brownian}.
In this case one can observe that the distributions of $\tau_+$ and $\tau_-$ are the same, thus denying the empirical evidence of gain/loss asymmetry.
However, it is well known that the empirical log-prices on market do not follow a Geometric Brownian Motion and an explicit form of the first hitting stopping time distribution is often not available for more complex stochastic processes.

In order to quantify the amount of difference between $\tau_+$ and $\tau_-$ we use the approach named \emph{Bayesian Estimation Supersedes the t-Test} (BEST) which is a bayesian generalization of the t-Test \cite{kruschke2013bayesian}. In fact, the trivial way to compare two distributions having a different mean and unequal size and variance is through the standard Welch test \cite{welch1947generalization}, a generalization of the t-Test. Nevertheless, this approach requires the strong assumption that both distributions are normal. It has been established in the literature that $\tau_{\pm}$ are not normal distributed, thus to address this problem we use the BEST procedure.

Bayesian estimation for two groups provides complete distributions of credible values for group means, standard deviations as well as the normality of the data. In the following we consider $x_{\pm} = \ln \tau_{\pm}$ to model the observations.
The basic assumption is that each group of dimension $N$ are represented by a Student-$t$ distribution with $\nu$ degrees of freedom, location parameter $\mu$ and scale parameter $\sigma$. In this setting the posterior probability conditional to the observations $D$ can be written as
\begin{equation}
    \underbrace{P(\theta|D)}_{posterior} = \underbrace{P(D|\theta)}_{likelihood} \times \underbrace{P(\theta)}_{prior} \times \underbrace{P^{-1}(D)}_{evidence} ,
\end{equation}
where $\theta = \{\mu_+, \sigma_+, \nu_+, \mu_-, \sigma_-, \nu_- \}$ are the parameters of the model, where the subscript $+$ and $-$ refers to $x_+$ and $x_-$ respectively. We assume the  data are independently sampled, therefore the likelihood is the multiplicative product across the data values of the probability density of the Student-$t$ distribution. The prior is the product of independent parameter distributions while the constant $P(D)$ is computed integrating the product of the likelihood and prior over the entire parameter space. The integral is impossible to compute analytically for several models and thus we use Markov Chain Monte Carlo (MCMC) to compute evidence and thus the full posterior distribution. For a reference on this topic see, for instance, \cite{gelman2013bayesian}.

In order to quantify the difference between two groups we use as a metric the \emph{effect size} introduced by the Cohen's $d$ \cite{cohen2013statistical}. This quantity is defined as the difference between the two means divided by a measure of the standard deviation for the data\footnote{This is a slightly different definition with respect to the one used in \cite{kruschke2013bayesian}, where the denominator is $s=\sqrt{(\sigma_+^2 + \sigma_-^2)/2}$. As discussed by Kruschke this is just a re-definition of the posterior distribution of $d$.}
\begin{equation}
\label{eq:effect_size}
    d = \frac{\mu_+ - \mu_-}{s}, \quad \mbox{where} \quad s= \sqrt{\frac{\sigma_+^2(N_+ -1) + \sigma_-^2(N_- -1)}{N_+ + N_- -2}}.
\end{equation}
We observe that the effect size is a number measuring the strength of the relationship between two variables in a statistical population.

\subsection{Hierarchical models}
In the original model proposed in \cite{kruschke2013bayesian} the data are distributed as a Student-$t$ whose probability distribution function (PDF) is 
\begin{equation}\label{t-distribution}
    f(x|\mu, \sigma, \nu) = \frac{\Gamma((\nu + 1)/2)}{\Gamma(\nu/2)} \left(\frac{1}{\pi \nu \sigma}\right)^{1/2} \left[ 1 + \frac{(x - \mu)^2}{\sigma \nu}\right] ^{-(\nu + 1)/2}
\end{equation}
where $\Gamma$ is the Gamma function. 

The pitfall of this approach is that the Student-$t$ distribution allows negative values, which are not feasible for $\tau_{\pm}$. In order to address this issue we also consider a different distribution for the data inspired by the first hitting time of a Geometrical Brownian Motion, i.e. an Inverse Gamma distribution, whose PDF is given by the following equation
\begin{equation}\label{igamma-distribution}
    f(x|\alpha, \beta) = \frac{\beta^\alpha}{\Gamma(\alpha)} x^{-\alpha - 1}\exp\left(-\frac{\beta}{x}\right)
\end{equation}
where the shape parameter $\alpha$ and the scale parameter $\beta$ are related to the mean $M$ of the distribution and the variance $S^2$  by the  following relationships
\begin{equation}  
M = \frac{\beta}{\alpha - 1}, \qquad S^2 = \frac{\beta^2}{(\alpha - 1)^{2}(\alpha - 2)}.
\end{equation}

Thus, in our analysis, we use two model to characterize the effect size $d$: the Student model \eqref{t-distribution}, which is the standard benchmark, and the distribution \eqref{igamma-distribution}, which is expected to be an improvement due to the positiveness of the distribution. In the following, we will address how to select the best model using a quantitative approach.

To better formalize the model in the Bayesian setting we have to define the priors for the model parameters. 

\subsubsection*{Student-$t$}
We use the following priors for the location parameter $\mu$, dispersion parameter $\sigma$ and degree of freedom $\nu$ of the Student distribution defined in Eq. \eqref{t-distribution}
\begin{align}
\label{eq:prior_student}
    \mu_{\pm} \sim \mathcal{N}(m_{\pm}, s^2_{\pm}) \qquad \sigma_{\pm} \sim \mathcal{U}(\sigma_{\rm low}, \sigma_{\rm high}) \qquad \nu_{\pm} \sim \mathcal{E}(t_{\pm}; q_{\pm}) 
\end{align}
where $\mathcal{N}$ is the normal distribution, $\mathcal{U}$ the uniform distribution and $\mathcal{E}$ the shifted exponential distribution whose PDF is $p(t; q) \propto \exp(t - q)$. In \eqref{eq:prior_student} we indicate as $m_{\pm}$ the empirical mean of the $x_{\pm}$ distribution, while $s^2_{\pm}$ is its empirical variance. The allowed interval for the dispersion prior is given by $\sigma_{\rm low} = 1$, $\sigma_{\rm high} = 100$. For the shifted exponential distribution we set the same parameters as \cite{kruschke2013bayesian}: $t_{\pm} = 1/29$ and $q_{\pm} = 1$. Notice that the parameter $\nu_{\pm}$ controls the normality of the distribution, the greater the value, the higher the normality of $x_{\pm}$.

\subsubsection*{Inverse Gamma (First hitting time)}
In this case we set the priors using the mean $M = \beta(\alpha - 1)^{-1}$ and variance $S^2 = \beta^2(\alpha - 1)^{-2}(\alpha - 2)^{-1}$ of the Inverse Gamma distribution \eqref{igamma-distribution}. In particular, we assume
\begin{equation}
    \label{eq:prior_fht}
    M_{\pm} \sim \mathcal{N}(m_{\pm}, s^2_{\pm}) \qquad S_{\pm} \sim \mathcal{U}(\sigma_{\rm low}, \sigma_{\rm high})
\end{equation}
where we use the same notation and settings as \eqref{eq:prior_student}. Using the relations between $M, S^2$ and $\alpha, \beta$ we can obtain the priors for the hierarchical model.

\subsection{Numerical implementation}
In this Section we discuss the numerical implementation of the performed analysis. We use the \texttt{python} package \texttt{PyMC3} \cite{Salvatier2016}. \texttt{PyMC3} is a probabilistic object-oriented programming developed to perform complex MCMC analysis using \texttt{Theano} \cite{bergstra2010theano} as a beck-end for efficient \texttt{C} evaluation. The library uses the so-called  No-U-Turn Sampler (NUTS) sampling algorithms \cite{hoffman2014no}, a variant of Hamiltonian Monte Carlo (HMC) \cite{duane1987hybrid}, see also \cite{neal2011mcmc, betancourt2017conceptual} for an introduction about this topic. Those samplers work well in high dimensions and within complex posterior distributions and allows complex models to be fit without knowledge about fitting algorithms. HMC and NUTS take advantage of gradient information from the likelihood to achieve much faster convergence than traditional sampling methods. In particular, Hamiltonian dynamics is used to make proposals for the Metropolis algorithm, thereby avoiding the slow exploration of the state space due to the diffusive behavior of simple random-walk. In other words, Hamiltonian dynamics can be used in several problems with continuous state spaces by simply introducing fictitious \emph{momentum} variables $p$. The Hamiltonian $\mathcal{H}$ is an energy function for the joint state of position $q$, and momentum $p$, and so defines a joint distribution for them, as follows: 
\begin{equation}
\label{eq:hamiltonian_proposal}
    P(q, p) = \frac{1}{\mathcal{Z}} \exp\left(-\frac{\mathcal{H}(q, p)}{T}\right)
\end{equation}
where $T$ is the temperature of the system and $\mathcal{Z}$ is the normalizing constant needed for this function to sum or integrate to one. The Hamiltonian can be decomposed as the sum of two terms, the potential $U(q)$ and the kinetic energy $K(q, p)$, thus \eqref{eq:hamiltonian_proposal} can be written as
\begin{equation}
\label{eq:hamiltonian_proposal_decomposed}
    P(q, p) = \frac{1}{\mathcal{Z}} \exp\left(-\frac{U(q)}{T}\right) \exp\left(-\frac{K(q, p)}{T}\right).
\end{equation}
In Bayesian statistics, the posterior distribution for the model parameters is the usual distribution of interest, and therefore these parameters will take the role of the position $q$. We can express the posterior distribution as a canonical distribution (fixing the temperature to be $T = 1$) using a potential energy function defined as follows
\begin{equation}
    U(q) =- \log \left[P(\theta) P(D|\theta) \right]
\end{equation}
where $P(\theta)$ is the prior density, and $P(D|\theta)$ is the likelihood function given data $D$.

In our setting we compute $n_{\rm chains} = 4$ Markov chains with $n_{\rm draw} = 4000$ draw each one. We also set $n_{\rm tune} = 2000$ as tune number to avoid \emph{cold start}, i.e. the first 2000 samplings are discarded to avoid issues with the starting point of the chain. We check the convergence of the chain using several techniques, in particular, we compute the Gelman-Rubin statistic, that checks MCMC convergence by analyzing the difference among Markov chains. The convergence is assessed by comparing the estimated between-chains and within-chain variances for each model parameter. Large differences between these variances indicate non-convergence \cite{gelman1992inference, brooks1998general}, see Appendix \ref{sec:appendix_diagnostics} for further details.

The goodness of MCMC has been estimated using the Effective Sample Size (ESS) statistics, which is a measure of how the samplings of the Markov chain are autocorrelated \cite{gelman2013bayesian, vehtari2020rank}. The higher the value the better the sampling procedure, see Appendix \ref{sec:appendix_diagnostics} for the definition.

To compare the hierarchical models we use the  Widely Applicable Information Criterion (WAIC) \cite{watanabe2010asymptotic, vehtari2017practical}. It is the generalized version of the Akaike Information Criterion (AIC) onto singular statistical models. WAIC estimates the effective number of parameters $p_{{\rm WAIC}}$ to adjust for overfitting. The parameter is defined as ${\rm WAIC} = -2({\rm lppd} - p_{{\rm WAIC}})$ where ${\rm lppd}$ is the log pointwise predictive density \cite{gelman2013bayesian}. The selected model is the one with the lower value of WAIC. 

\section{Dataset}
\label{sec:dataset}
In our analysis we consider the historical daily prices of the following relevant financial indices:
    \begin{itemize}
    \item Australia's leading share market index (ASX200), based on the 200 largest stocks listed on the Australian Securities Exchange. 
    \item The Bovespa Index (BVSP), the benchmark index of about 70 stocks that are traded on the Brazilian stock market.
    \item The DAX stock market index, consisting of the 30 major German companies trading on the Frankfurt Stock Exchange.
    \item The Dow Jones Industrial Average (DJI30), which measures the stock performance of 30 large companies in the USA Stock Exchange.
    \item The EURO STOXX 50 (EUROSTOXX50), the reference index for the Eurozone.
    \item The Hang Seng Index (HSI), the reference market-capitalization-weighted stock-market index in Hong Kong composed of the 50 largest companies in the market.
    \item  The Nikkei Stock Average (NIKKEI225), the reference stock market index for the Tokyo Stock Exchange.
    \item  The Standard \& Poors 500 (S\&P500), a reference stock market index measuring the performance of 500 largest companies listed on stock exchanges in the USA. 
    \item The  SSE Composite Index (SHANGAI), the reference index for the Stock Exchange in Shangai.
    \item The Chicago Board Options Exchange's CBOE Volatility Index (VIX), the reference measure of the stock market's expectation of volatility based on S\&P 500 options.
    \end{itemize}
All considered time series cover the period from January 01, 2008, to December 31, 2020. We refer to Table \ref{tab:summary_indices} for a summary of the basic statistical properties of our dataset: we indicate statistics for the raw data as well as those for the filtered time-series, where we set the filter size $f=252$. In the considered dataset there exists a correlation about the return time-series. As an example, considering the raw returns, i.e. not detrended, the correlation between DJI30 and S\&P500 is 0.97, while the one between BVSP and DAX is 0.63. Hence, we expect similar behavior among similar instruments.

\begin{table}[h!]
    \centering
    
    \begin{tabular}{c ccc | ccc}
    \toprule
    \toprule
     & \multicolumn{3}{c}{ Raw } & \multicolumn{3}{c}{ Filtered } \\
    {\bf Index} &  {\bf Counts} &  {\bf Mean} &  {\bf Std} &  {\bf Counts} &  {\bf Mean} &  {\bf Std} \\
    \midrule
    ASX200      &   3340 &   8.55 &  0.15 &           3089 &         -0.01 &         0.04 \\
    BVSP        &   3216 &  11.06 &  0.26 &           2965 &         -0.01 &         0.08 \\
    DAX         &   3302 &   9.07 &  0.32 &           3051 &         -0.01 &         0.06 \\
    DJI30       &   3273 &   9.69 &  0.36 &           3022 &         -0.01 &         0.04 \\
    EUROSTOXX50 &   2407 &   8.03 &  0.14 &           2156 &         -0.00 &         0.05 \\
    HSI         &   3221 &  10.04 &  0.17 &           2970 &         -0.01 &         0.06 \\
    NIKKEI225   &   3206 &   9.59 &  0.36 &           2955 &         -0.00 &         0.06 \\
    S\&P500      &   3274 &   7.49 &  0.39 &           3023 &         -0.01 &         0.04 \\
    SHANGAI     &   3165 &   7.93 &  0.20 &           2914 &         -0.00 &         0.07 \\
    VIX         &   3278 &   2.92 &  0.40 &           3027 &          0.05 &         0.22 \\
    \bottomrule
    \bottomrule
    \end{tabular}
    
    \caption{\small{Summary of the data used in the analysis: the number of observations, the mean values and the standard deviations are reported for both the raw data and for the filtered ones.}}
    \label{tab:summary_indices}
\end{table}

\begin{figure}[h!]
    \centering
    \includegraphics[width=1\textwidth]{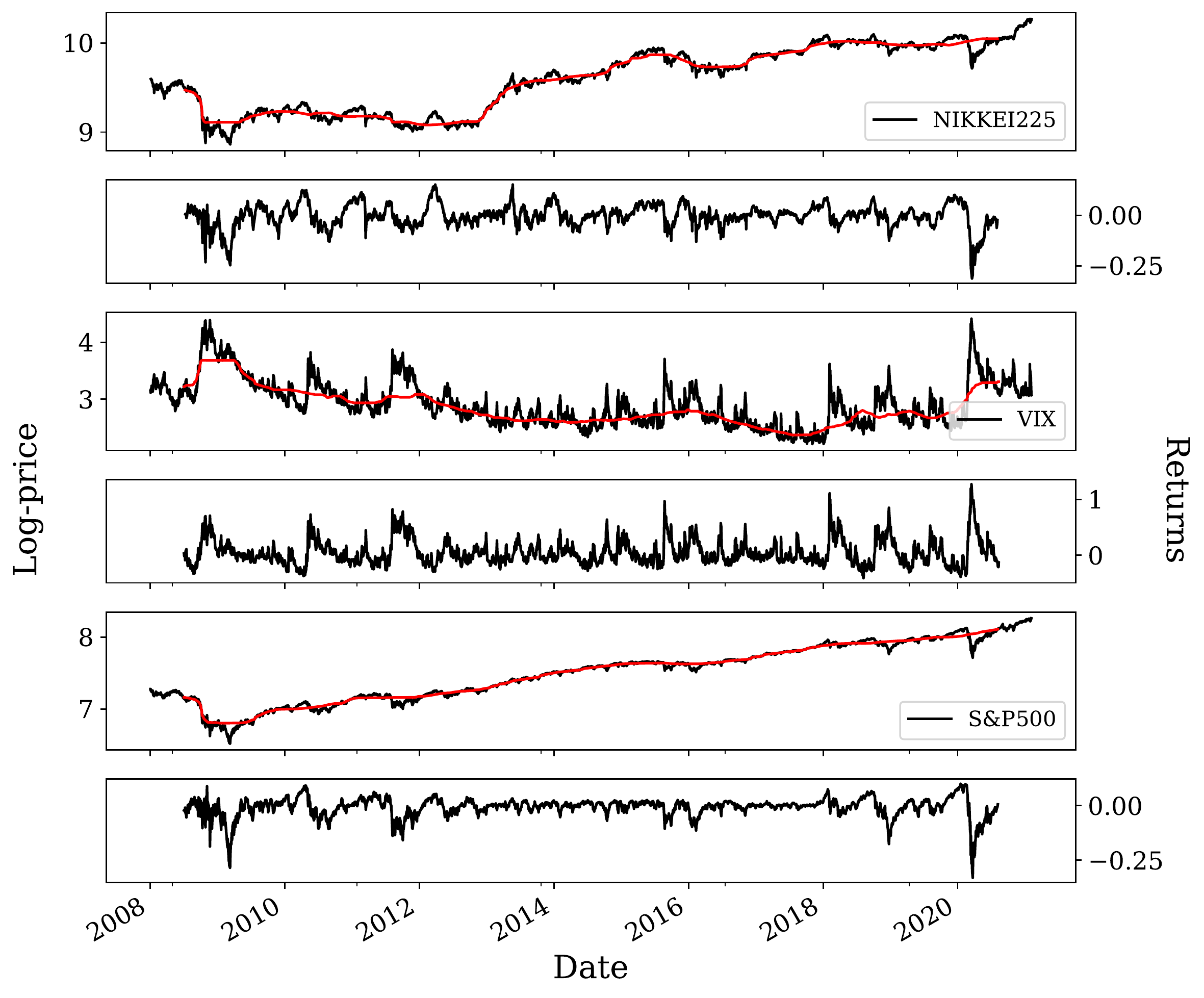}
    \caption{   \small{Log prices and filtered time-series for few selected indices: NIKKEI225 (lower panels), VIX (middle panels) and S\&P500 (lower panels) as a function of the trading calendar. In red are shown the rolling median trends for a filter size $f=252$.}}
    \label{fig:trend}
\end{figure}

We remark that in the considered basket of indices VIX has a different meaning with respect to others because it is related to the implied volatility of S\&P500 options \cite{whaley2009understanding}. As a consequence, the returns $r_t$ dynamic looks different with respect to other time-series, and the raw returns are negatively correlated at -0.75 with S\&P500. To better illustrate this behavior, in Figure \ref{fig:trend} the raw log price $S_t$, as well as the filtered signal of few selected indices, are shown: NIKKEI225 (upper panels), VIX (middle panels) and S\&P500 (lower panels). It is straightforward to observe that for equities' indices the filtered time series follows some kind of random walk while in the case of VIX mainly positive spikes are present.

\begin{figure}[h!]
    \centering
    
    \centering
    \includegraphics[width=.9\textwidth]{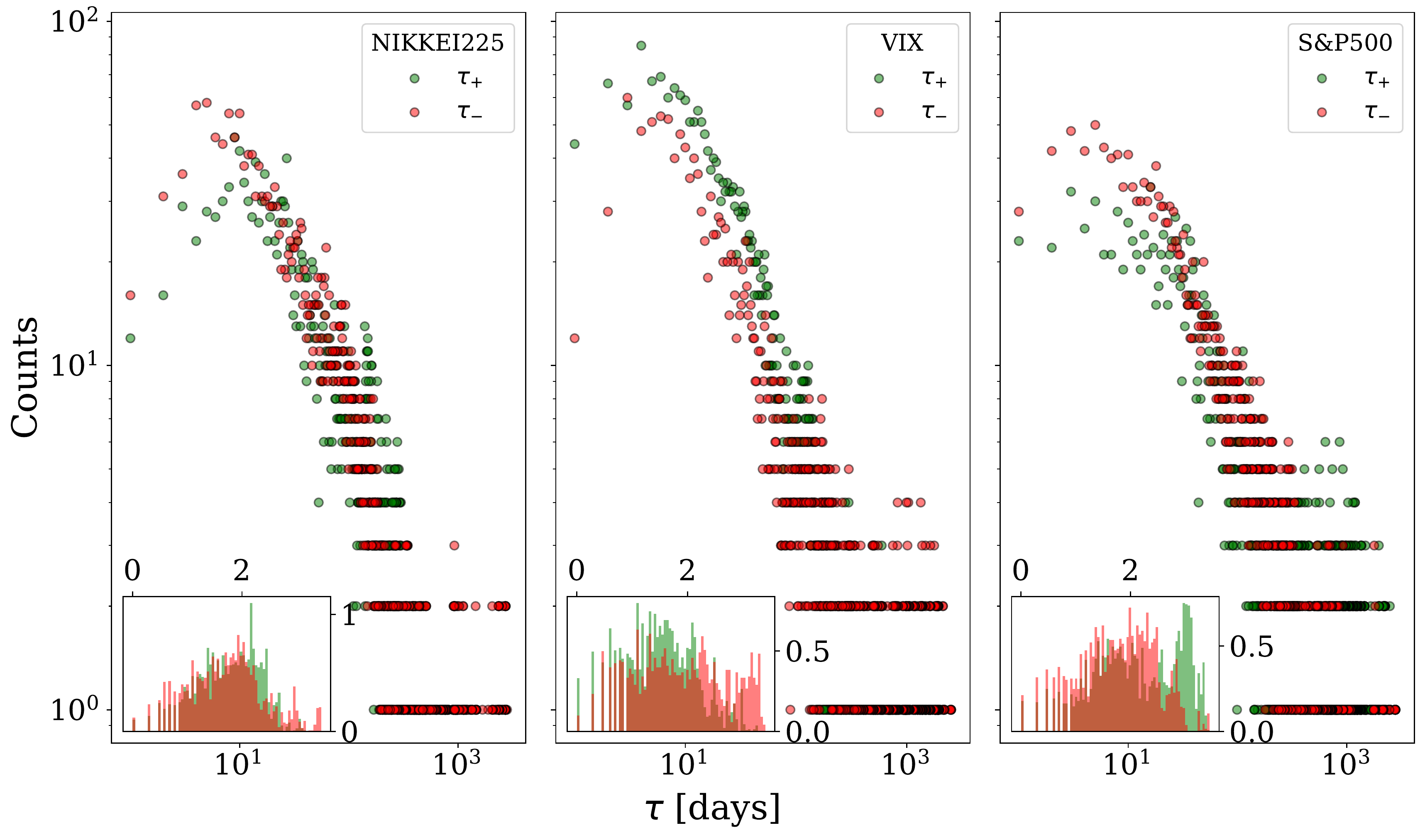}

    \caption{\small{Distribution of the first hitting time as defined in Eq. \eqref{eq:tau_definition} for few indices, NIKKEI225 (left panel), VIX (middle panel) and S\&P500 (right panel), in green $\tau_+$ while in red $\tau_-$. In the inset the density as a function of $\log_{10} \tau_{\pm}$. }}
    \label{fig:counts_histograms}
\end{figure}

The asymmetry parameters $\tau_{\pm}$ introduced in Eq. \eqref{eq:tau_definition}, quantifies the amount of difference between the \emph{velocity} of up-down and drawdown movements. Empirically, we observe a common pattern for all the indices but VIX. The distribution of $\tau_+$ has a larger value for the peak, thus we expect $d \gtrsim 0$, while in the case of VIX it seems that $\tau_-$ has this behavior. This can be observed in Figure \ref{fig:counts_histograms}, where we display the distributions of $\tau_{\pm}$ for the same instruments as Figure \ref{fig:trend}. 

\section{Results}
\label{sec:section-results}

In the following, we discuss the main results of the study of the effect size parameter $d$ defined in Eq. \eqref{eq:effect_size}. Here the filter size of $f=252$ days has been used and $\rho$ has been fixed to be the sample standard deviation. We compare the two models introduced in Section \ref{sec:methodology}, namely the Student-$t$ and the Inverse Gamma proposal for the likelihood. The results are reported in Table \ref{tab:analysis_student_t} for the Student-$t$ distribution while in Table \ref{tab:analysis_fht} are summarized those of Inverse Gamma.

\begin{table}[h!]
    \centering
    \begin{tabular}{ccccc}
\toprule
\toprule
      {\bf Index} &    $\rho$ &   $d$       &  ESS &       WAIC  \\
\midrule
      ASX200 &  0.039 &  0.259 $\pm$  0.027 &    4656 & -11007.340 $\pm$   47.142 \\
        BVSP &  0.078 &  0.041 $\pm$  0.029 &    5779 &  -9215.047 $\pm$   50.235 \\
         DAX &  0.058 &  0.303 $\pm$  0.029 &    5555 &  -9903.044 $\pm$   47.971 \\
       DJI30 &  0.043 &  0.461 $\pm$  0.028 &    6581 & -10982.727 $\pm$   44.261 \\
 EUROSTOXX50 &  0.050 &  0.373 $\pm$  0.035 &    5990 &  -7254.869 $\pm$   37.863 \\
         HSI &  0.056 &  0.135 $\pm$  0.029 &    6802 &  -9649.320 $\pm$   42.360 \\
   NIKKEI225 &  0.056 &  0.066 $\pm$  0.029 &    5028 &  -8984.813 $\pm$   49.780 \\
      S\&P500 &  0.044 &  0.477 $\pm$  0.028 &    5990 & -10972.168 $\pm$   44.446 \\
     SHANGAI &  0.071 &  0.161 $\pm$  0.030 &    5804 &  -8971.086 $\pm$   43.375 \\
         VIX &  0.223 & -0.543 $\pm$  0.030 &    5238 & -10134.544 $\pm$   43.957 \\
\bottomrule
\bottomrule
\end{tabular}
    \caption{\small{Results for the posterior mean of effect size $d$ and its standard deviation using the Student-$t$ model. For each index the $\rho$ level is also indicated. Additionally, the effective sample size ESS discussed in Appendix \ref{sec:appendix_diagnostics}, as well as the WAIC parameters and standard deviations, are reported.}}
    \label{tab:analysis_student_t}
\end{table}

\begin{table}[h!]
    \centering
    \begin{tabular}{ccccc}
\toprule
\toprule
      {\bf Index} &    $\rho$ &   $d$       &  ESS &       WAIC  \\
\midrule
      ASX200 &  0.039 &  0.159 $\pm$  0.026 &    1957 & -11813.257 $\pm$   76.034 \\
        BVSP &  0.078 &  0.013 $\pm$  0.027 &    2444 & -10088.991 $\pm$   89.921 \\
         DAX &  0.058 &  0.173 $\pm$  0.027 &    2164 & -10786.804 $\pm$   76.233 \\
       DJI30 &  0.043 &  0.253 $\pm$  0.026 &    2063 & -12224.544 $\pm$   82.602 \\
 EUROSTOXX50 &  0.050 &  0.225 $\pm$  0.031 &    1971 &  -7782.176 $\pm$   54.619 \\
         HSI &  0.056 &  0.084 $\pm$  0.027 &    2201 & -10401.920 $\pm$   70.030 \\
   NIKKEI225 &  0.056 &  0.001 $\pm$  0.026 &    2060 &  -9781.575 $\pm$   80.745 \\
      S\&P500 &  0.044 &  0.261 $\pm$  0.026 &    1932 & -12260.829 $\pm$   81.735 \\
     SHANGAI &  0.071 &  0.073 $\pm$  0.028 &    2155 &  -9878.896 $\pm$   72.530 \\
         VIX &  0.223 & -0.307 $\pm$  0.025 &    1950 & -10834.394 $\pm$   62.396 \\
\bottomrule
\bottomrule
\end{tabular}
    \caption{\small{The same as Table \ref{tab:analysis_student_t} but in the case of Inverse Gamma hierarchical model.}}
    \label{tab:analysis_fht}
\end{table}

\begin{figure}[ht!]
    \centering
    \includegraphics[width=0.9\textwidth]{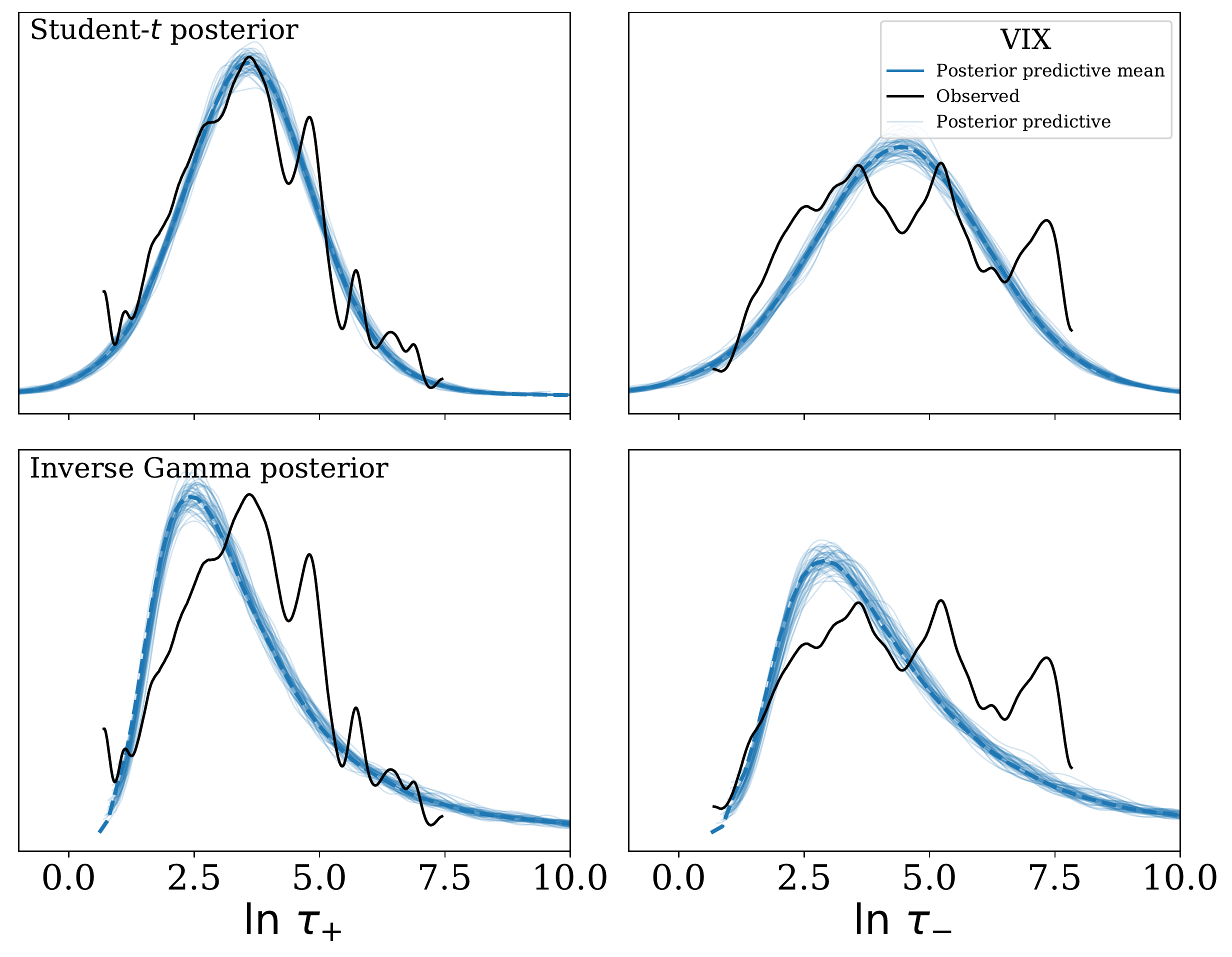}
    \caption{\small Posterior densities of $\ln \tau_{\pm}$ for the VIX index assuming Student-$t$ (upper panels) and Inverse Gamma (lower panels). The black solid line is a kernel smooth estimate of the observations, the shadow blue lines are samples extracted from the posterior distributions, while the tick dashed blue line is the posterior mean.}
    \label{fig:fit_VIX}
\end{figure}

For both models the posterior distributions the effect size $d$ is a positive value except for VIX, thus confirming the empirical evidence that the posterior mean of $\tau_+$ is greater than those of $\tau_-$. As a consequence, the hitting time of up-down movements is slower than those of drawdowns, as we aimed to prove. To better illustrate how the hierarchical models fit the realization of $\tau_{\pm}$ we show in Figure \ref{fig:fit_VIX} the results for the posterior densities in the case of the VIX index. In the panels are reported the kernel density estimate \cite{scott2015multivariate} of the observed quantities (black solid line), a random sampling of the posterior (shadow lines) and the posterior mean (thick dashed line).

Observing the plots few remarks are relevant: the observations do not present a smooth curve, in fact, several peaks are present, hence the proposed hierarchical models can capture only the coarse shape of the distributions. The Student-$t$ model is generally capable to fit the peak of the distribution, but it has a domain in the not-allowed region $\tau_{\pm} < 0$, therefore the WAIC value is higher than the other model. On the other hand, the Inverse Gamma model (lower panels) is able to capture the right tail of the distribution but is less capable to fit the left tail. The magnitude of the effect size $d$, as reported in Tables \ref{tab:analysis_student_t} and \ref{tab:analysis_fht}, is lower for the Inverse Gamma model, this is due to the fact that the posterior standard deviation is broad in this model. In fact, we get the following mean of the posterior for the parameters defined in Eq. \eqref{eq:effect_size}: $\mu_+ = 3.58$ $(3.79)$, $\mu_- = 4.40$ $(4.65)$, $\sigma_+ = 1.28$ $(2.41)$ and $\sigma_- = 1.73$ $(3.73)$ for the Student-$t$ (Inverse Gamma) hierarchical model. Similar considerations are possible for all the indices.

\begin{figure}
    \centering
    \includegraphics[width=0.9\textwidth]{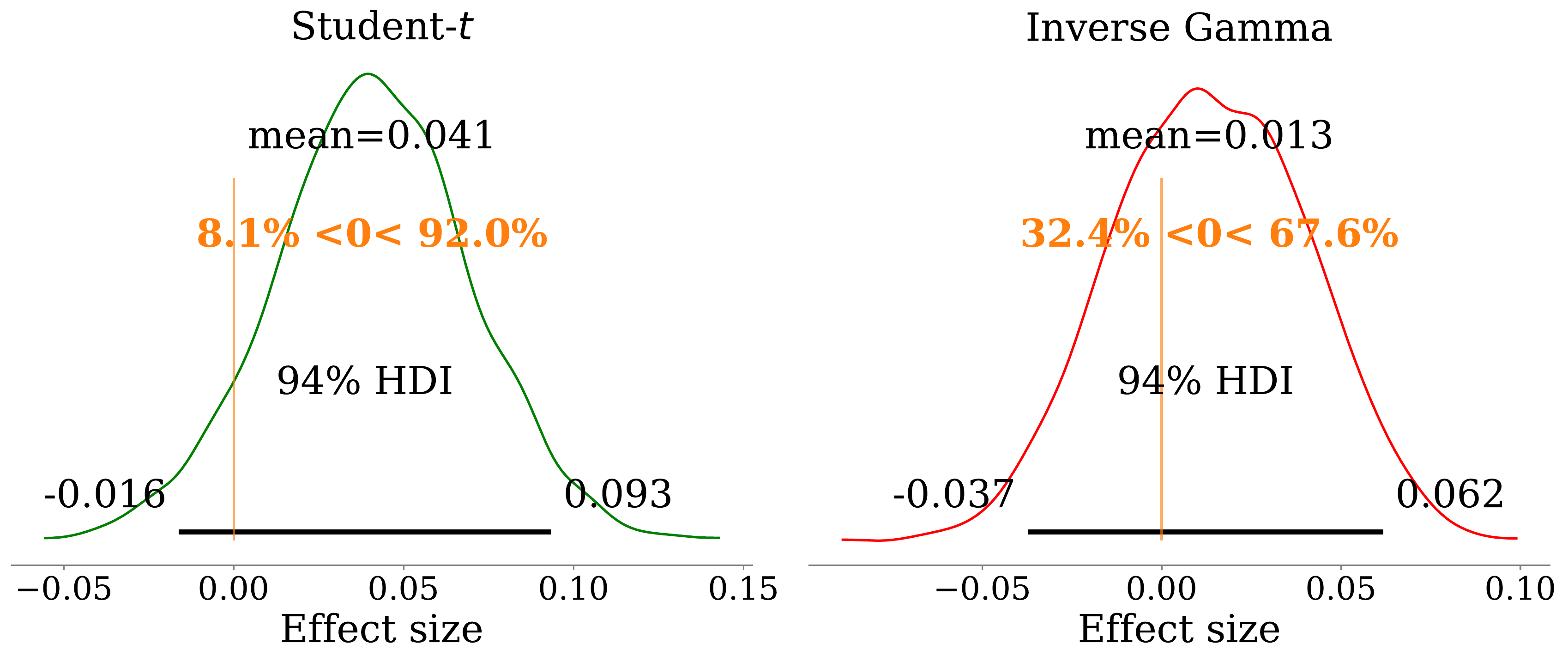}
    \caption{\small Posterior densities distribution of the effect size $d$ defined in Eq. \eqref{eq:effect_size} for the Brazilian index BVSP in the Student-$t$ model (left panel) and Inverse Gamma one (right panel). In the plots are indicated the mean value and the 94\% HDI with a thick horizontal line. The vertical orange line is the reference value, $d=0$ and the percentages are the probability to be lower or greater than the reference value.}
    \label{fig:posterior_bvsp}
\end{figure}

From Tables \ref{tab:analysis_student_t} and \ref{tab:analysis_fht} it is worth noting that for BVSP and NIKKEI225, $d$ is close to zero in both the models, thus our belief about the existence of a gain/loss asymmetry for these indices is less robust. Indeed, the Highest Density Interval (HDI) at 94\% of probability is $[-0.016, 0.093]$ ($[-0.037, 0.062]$) for the former and $[0.010, 0.118]$ ($[-0.048,	0.050]$) for the latter using the Student-$t$ (Inverse Gamma) model. In order to better illustrate the results, the posterior densities for the effect size $d$ in the case of BVSP index for both hierarchical models are shown in Figure \ref{fig:posterior_bvsp}. We observe a mono-modal distribution, thus the mean is a safe punctual estimation of the posterior distribution. It is also possible to note that the posterior has a bell-like shape, so we use the standard deviation to quantify the uncertainty in the previous tables. In the figure are also reported the HDI and the reference value $d=0$, which indicates no asymmetry. We observe that in the case of Inverse Gamma distribution it exists a probability of $32.4\%$ to have $d < 0$. A similar analysis can be performed for any indices.

As expected, in Tables \ref{tab:analysis_student_t} and\ref{tab:analysis_fht} the case of VIX index the $d$ posterior mean is lower than zero and the HDI are $[-0.598, -0.488]$ and $[-0.354, -0.260]$ respectively for Student-$t$ and Inverse Gamma model, hence the updraws movements are faster. This is related to the fact that VIX is often indicate as the \emph{investor fear gauge} in the equity market \cite{whaley2000investor}.

In general, it is possible to observe that WAIC is smaller in the case of the Inverse Gamma distribution model with respect to the Student-$t$. Thus, from a model selection perspective, we can consider the former as the best model to describe the observed data. This is expected since the Student-$t$ distribution allows for negative values of hitting time, while the Inverse Gamma distribution is positive definite.

Finally, the MCMC convergence has been investigated. In our analysis he Gelman-Rubin statistic $R_c$, discussed in Appendix \ref{sec:appendix_diagnostics}, is close to one for all the fit performed, thus we neglected the values in Table \ref{tab:analysis_student_t} and \ref{tab:analysis_fht}. On the other hand, we observe that the ESS is larger in the case of the Student-$t$ model for all the indices considered. As a consequence, the sampling procedure is more efficient in the case of Student-$t$ model.

\subsection{Sensitivity analysis} 

In this Section, we quantify the impact of the additional parameters used in the model, namely the filter size $f$ and the threshold level for first passage $\rho$. The analysis performed above is sensible to these parameters and our findings should consider the impact of these values.

\subsubsection{Filter size $f$ }
\label{sec:filter_size_scan}
For the filter size $f$ we perform a scan in the range
$$
f \in \{150, 175, 200, 225, 250, 275, 300, 325\}$$
where $f$ is measured in business days. A similar analysis has been performed in Ref. \cite{liu2020analysis}, where the authors analyze the behavior of the market to investigate the rate of change in positive and negative price movements as a function of the time span.  We compute the posterior distribution of the effect size $d$ for both the proposed hierarchical models discussed in Section \ref{sec:methodology}. The parameter $\rho$ is fixed to the same values of the previous analysis, see Table \ref{tab:analysis_student_t}. The mean values of the posterior distribution of the effect size $d$ are reported in Table \ref{tab:sensitivity_filter_size_student_t} for the Student-$t$ hierarchical model and in Table \ref{tab:sensitivity_filter_size_fht} for the Inverse Gamma one. We perform a new numerical scan, thus the obtained results for $f=250$ are slightly different. This can be used as a useful cross-check to assess the MCMC convergence.

\begin{table}[ht!]
    \centering
    
\begin{tabular}{ccccccccc}
\toprule
\toprule
$f$ [days] &    150 &    175 &    200 &    225 &    250 &    275 &    300 &    325 \\

\midrule
ASX200      &  0.397 &  0.291 &  0.271 &  0.239 &  0.254 &  0.259 &  0.225 &  0.209 \\
BVSP        &  0.187 &  0.168 &  0.094 &  0.091 &  0.045 &  0.004 & -0.027 & -0.028 \\
DAX         &  0.352 &  0.360 &  0.355 &  0.314 &  0.293 &  0.283 &  0.284 &  0.263 \\
DJI30       &  0.453 &  0.470 &  0.483 &  0.464 &  0.451 &  0.448 &  0.445 &  0.441 \\
EUROSTOXX50 &  0.247 &  0.277 &  0.375 &  0.343 &  0.365 &  0.348 &  0.315 &  0.327 \\
HSI         &  0.140 &  0.100 &  0.113 &  0.094 &  0.117 &  0.156 &  0.147 &  0.115 \\
NIKKEI225   &  0.169 &  0.118 &  0.121 &  0.115 &  0.058 & -0.004 & -0.019 & -0.018 \\
S\&P500      &  0.500 &  0.536 &  0.530 &  0.511 &  0.483 &  0.474 &  0.492 &  0.477 \\
SHANGAI     &  0.104 &  0.144 &  0.162 &  0.146 &  0.159 &  0.128 &  0.100 &  0.114 \\
VIX         & -0.487 & -0.338 & -0.459 & -0.521 & -0.523 & -0.514 & -0.440 & -0.432 \\
\bottomrule
\bottomrule
\end{tabular}
    
    \caption{\small{Mean value of the posterior distribution of $d$ as a function of the filter size $f$ using the Student-$t$ hierarchical model.}}
    \label{tab:sensitivity_filter_size_student_t}
\end{table}

\begin{table}[ht!]
    \centering
    \begin{tabular}{ccccccccc}
\toprule
\toprule
$f$ [days] &    150 &    175 &    200 &    225 &    250 &    275 &    300 &    325 \\

\midrule
ASX200      &  0.247 &  0.187 &  0.169 &  0.151 &  0.155 &  0.159 &  0.133 &  0.124 \\
BVSP        &  0.126 &  0.109 &  0.058 &  0.054 &  0.018 & -0.012 & -0.032 & -0.035 \\
DAX         &  0.212 &  0.212 &  0.205 &  0.179 &  0.165 &  0.160 &  0.160 &  0.144 \\
DJI30       &  0.246 &  0.262 &  0.267 &  0.258 &  0.249 &  0.243 &  0.241 &  0.237 \\
EUROSTOXX50 &  0.157 &  0.172 &  0.234 &  0.209 &  0.223 &  0.212 &  0.191 &  0.190 \\
HSI         &  0.094 &  0.061 &  0.064 &  0.055 &  0.073 &  0.098 &  0.091 &  0.076 \\
NIKKEI225   &  0.087 &  0.048 &  0.049 &  0.038 & -0.002 & -0.047 & -0.059 & -0.061 \\
S\&P500      &  0.266 &  0.288 &  0.288 &  0.278 &  0.263 &  0.256 &  0.264 &  0.255 \\
SHANGAI     &  0.038 &  0.061 &  0.072 &  0.063 &  0.068 &  0.050 &  0.035 &  0.050 \\
VIX         & -0.272 & -0.186 & -0.255 & -0.291 & -0.293 & -0.286 & -0.243 & -0.237 \\
\bottomrule
\bottomrule
\end{tabular}
    \caption{\small{The same as Table \ref{tab:sensitivity_filter_size_student_t} in the case of Inverse Gamma hierarchical model.} }
    \label{tab:sensitivity_filter_size_fht}
\end{table}

\begin{figure}[h!]
    \centering
    \includegraphics[width=0.45\textwidth]{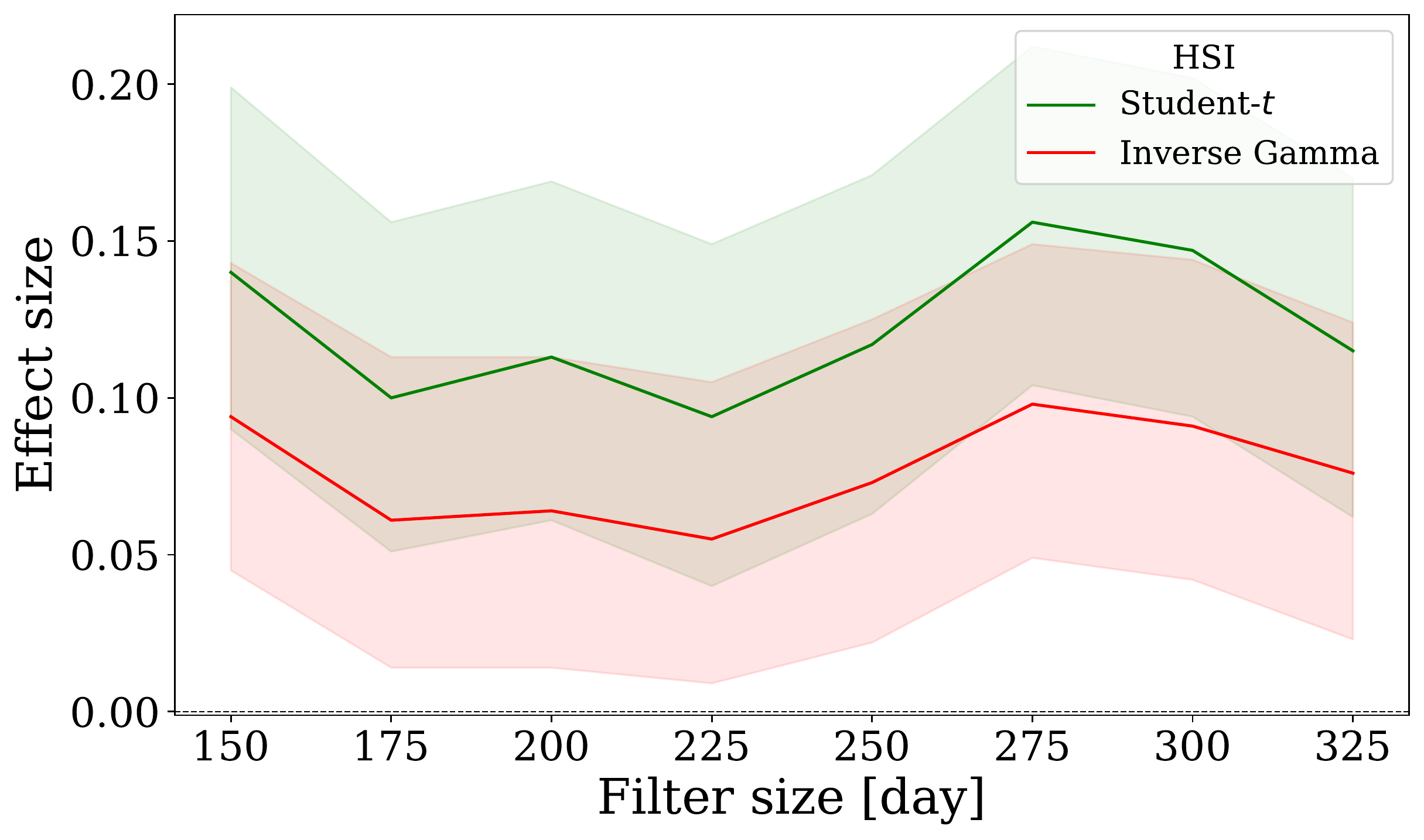}
    \includegraphics[width=0.45\textwidth]{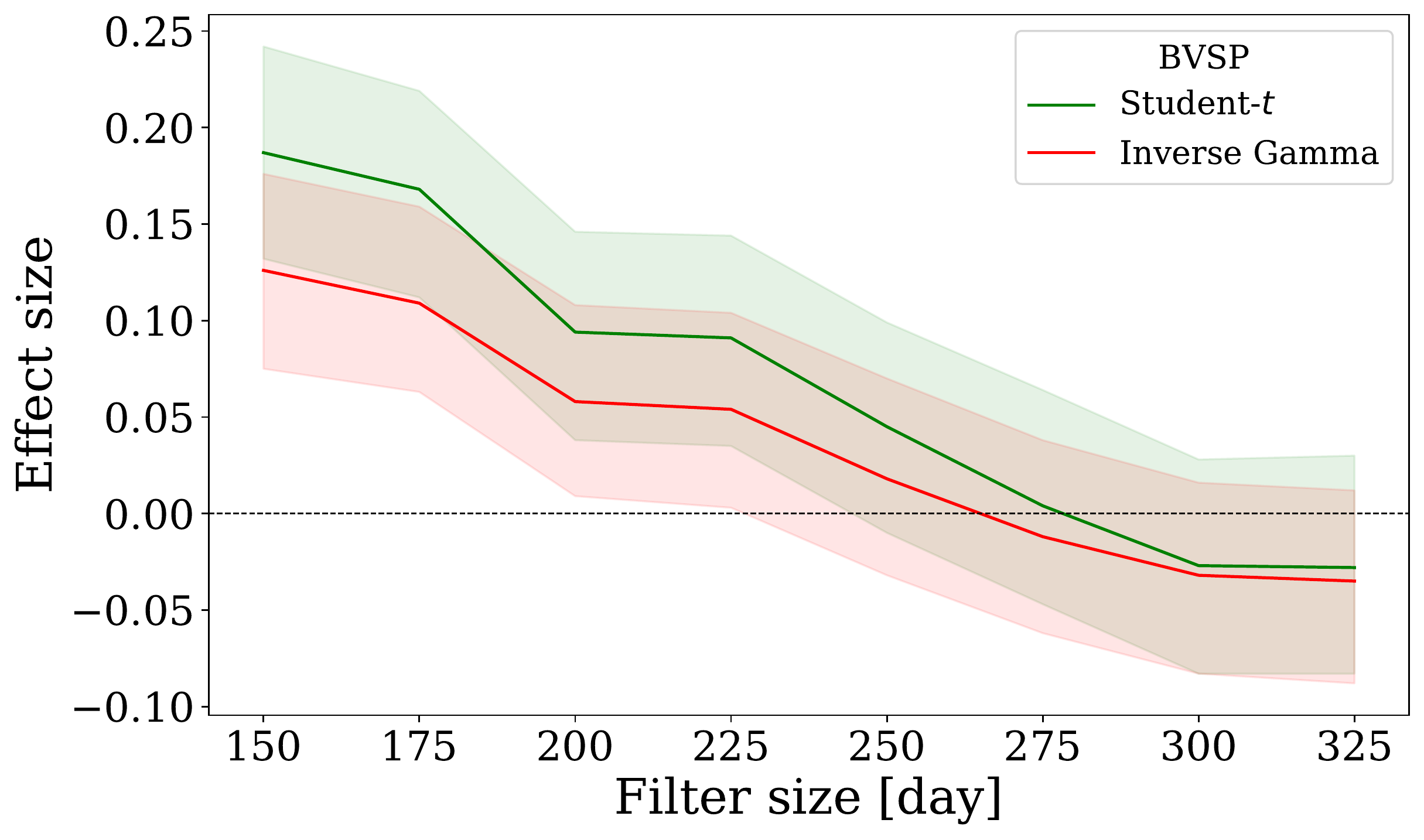}
    \caption{\small{Posterior distribution of the effect size $d$ defined in Eq. \eqref{eq:effect_size} as a function of the number of business days used as filter size $f$. In green the results for the Student-$t$ model while in red those of Inverse Gamma distribution. The thick lines are the posterior mean values, reported in Table \ref{tab:sensitivity_filter_size_fht} for Student-$t$ and Table \ref{tab:sensitivity_filter_size_fht} for Inverse Gamma. The shadow areas are the HDI at 94\%. In the left panel the Hong Kong index HSI while in the right panel the Brazilian BVSP. }}
    \label{fig:filter_size_scan}
\end{figure}

For the sake of clarity, we neglect in both tables the confidence level on the estimation of the mean value. The confidence intervals are roughly in the range $0.025 \div 0.035$ for all the mean values. To better illustrate the effect of $f$ on $d$ we show in Figure \ref{fig:filter_size_scan} the HDI at 94\% of the posterior distribution of $d$ (shadow areas) as well as the mean value (thick lines) for two interesting cases: HSI (left panel) and BVSP (right panel) assuming Student-$t$ (green) or Inverse Gamma (red) hierarchical model. In the case of HSI $d$ is almost stationary as a function of $d$ and our belief is that $d$ is always positive, while in the case of BVSP the posterior mean value is a decreasing function, and it is negative for $f \gtrsim 250$ business days. Nevertheless, the HDI has positive values also for large values of $f$.

As done in the previous Section we analyze which model is the most plausible using the WAIC methodology. For the sake of simplification, we do not report here the values. We found that for all the values of $f$ the Inverse Gamma hierarchical model is better than Student-$t$, thus this model is consistent for the different magnitude of the filter size. We also studied the convergence of the MCMC through the Gelman-Rubin statistic discussed in Appendix \ref{sec:appendix_diagnostics}, obtaining a value of $R_c \simeq 1.01$ for all the simulations, hence we can safely assume that the chains converge. We also check the ESS: we observe that the Student-$t$ model has a large value of ESS, meaning that the number of independent samples in the chains is greater in the case of Student-$t$ model. For the model based on the Inverse Gamma we get ${\rm ESS} \gtrsim 2000$, therefore the estimation error is roughly $(2000)^{-1/2} \lesssim 0.02$ which is sufficient in our analysis.

\subsubsection{Crossing level $\rho$} 
In order to study the sensitivity of the effect size with respect to the threshold crossing level $\rho$ we consider the sample standard deviation ${\rm Std}$ of each distribution as in Section \ref{sec:section-results} and we rescale this value, multiplying it by a series of fixed scale values. In particular, for each time series we scan the following  values:
$$
        \rho    \in {\rm Std} \times \{0.1, 0.25, 0.50, 0.75, 1, 1.25, 1.50, 1.75, 2, 2.25, 2.5\},
$$
while the filter size is fixed to $f = 252$ business days. The mean values of the posterior distribution of the effect size are  reported in Table \ref{tab:sensitivity_rho_student_t} for the Student-$t$ hierarchical model and in Table \ref{tab:sensitivity_rho_fht} for the Inverse
Gamma one. As in the case of filter size $f$ we perform a new numerical scan, thus the value obtained for $\rho/{\rm Std} = 1$ are different to those obtained at the beginning of Section \ref{sec:section-results}, however, the values are similar, thus we can safety assume the MCMC convergence.

\begin{table}[ht!]
    \centering\tiny{
\begin{tabular}{ccccccccccccc}
\toprule
\toprule
$\rho/{\rm Std}$ scale factor [\%]  &    10  &    25  &    50  &    75  &    100 &    125 &    150 &    175 &    200 &    225 &    250 \\
\midrule
ASX200      & -0.126 & -0.103 &  0.087 &  0.185 &  0.259 &  0.331 &  0.371 &  0.383 &  0.352 &  0.233 &  0.024 \\
BVSP        & -0.008 & -0.008 &  0.019 &  0.041 &  0.041 &  0.036 &  0.017 & -0.031 & -0.161 & -0.353 & -0.550 \\
DAX         & -0.153 & -0.010 &  0.102 &  0.227 &  0.304 &  0.311 &  0.265 &  0.147 &  0.010 & -0.225 & -0.511 \\
DJI30       & -0.224 & -0.035 &  0.225 &  0.358 &  0.461 &  0.464 &  0.488 &  0.520 &  0.544 &  0.626 &  0.440 \\
EUROSTOXX50 & -0.211 & -0.020 &  0.167 &  0.315 &  0.373 &  0.449 &  0.525 &  0.475 &  0.450 &  0.283 & -0.059 \\
HSI         & -0.095 &  0.012 &  0.054 &  0.094 &  0.134 &  0.201 &  0.256 &  0.295 &  0.324 &  0.311 &  0.303 \\
NIKKEI225   & -0.126 & -0.053 &  0.071 &  0.090 &  0.064 & -0.010 & -0.104 & -0.253 & -0.426 & -0.590 & -0.756 \\
S\&P500      & -0.268 &  0.004 &  0.279 &  0.422 &  0.477 &  0.470 &  0.499 &  0.603 &  0.617 &  0.531 &  0.279 \\
SHANGAI     & -0.068 & -0.061 &  0.014 &  0.113 &  0.161 &  0.144 &  0.136 &  0.148 &  0.147 &  0.182 &  0.277 \\
VIX         &  0.147 &  0.061 & -0.219 & -0.427 & -0.543 & -0.606 & -0.585 & -0.472 & -0.362 & -0.217 & -0.062 \\
\bottomrule
\bottomrule
\end{tabular}
    \caption{\small{Mean value of the posterior distribution of $d$ as a function of the crossing level $ \rho$ using the Student-$t$ hierarchical model.}}
    \label{tab:sensitivity_rho_student_t}}
\end{table}

\begin{figure}
    \centering
    \includegraphics[width=0.45\textwidth]{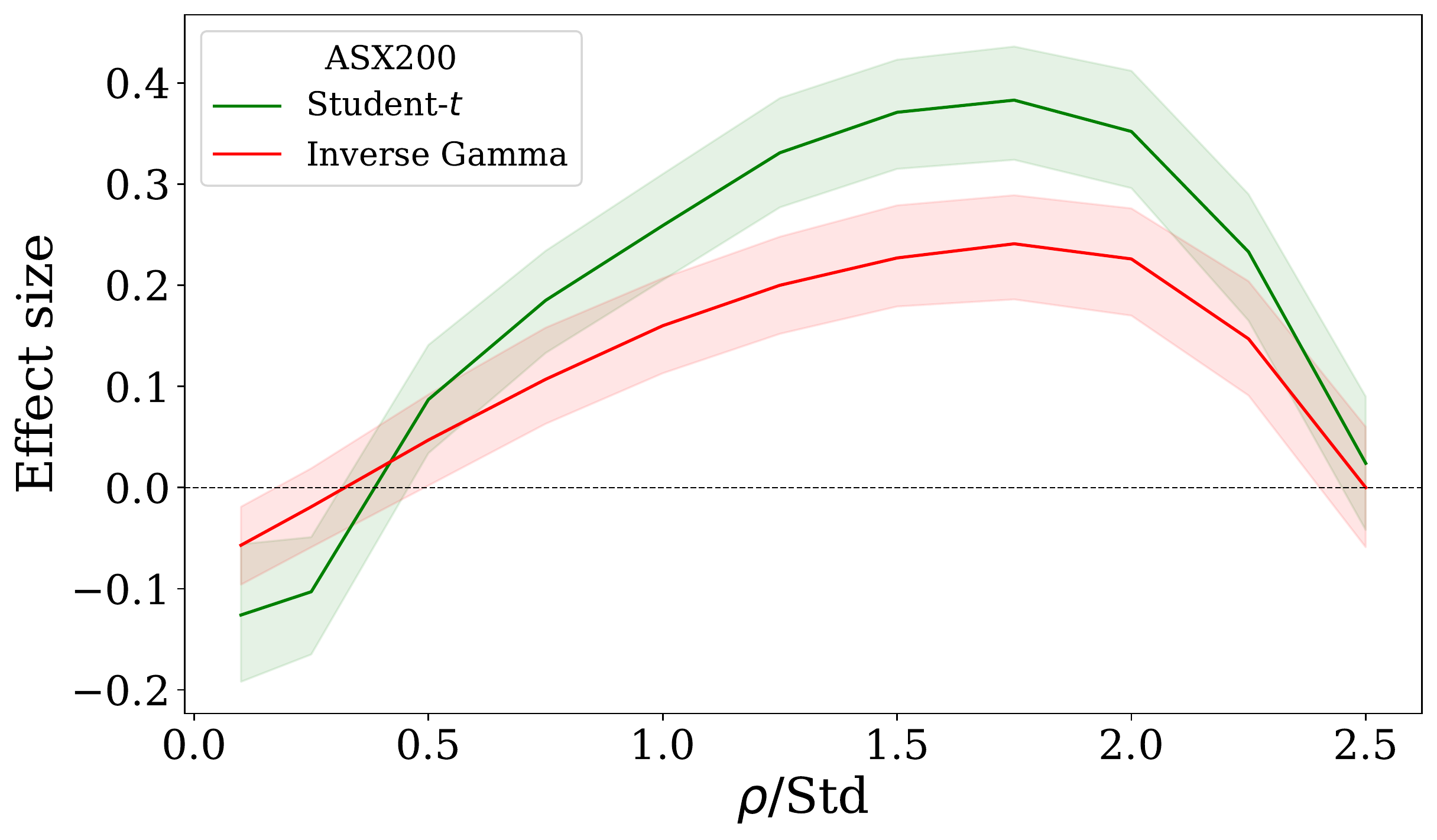}
    \includegraphics[width=0.45\textwidth]{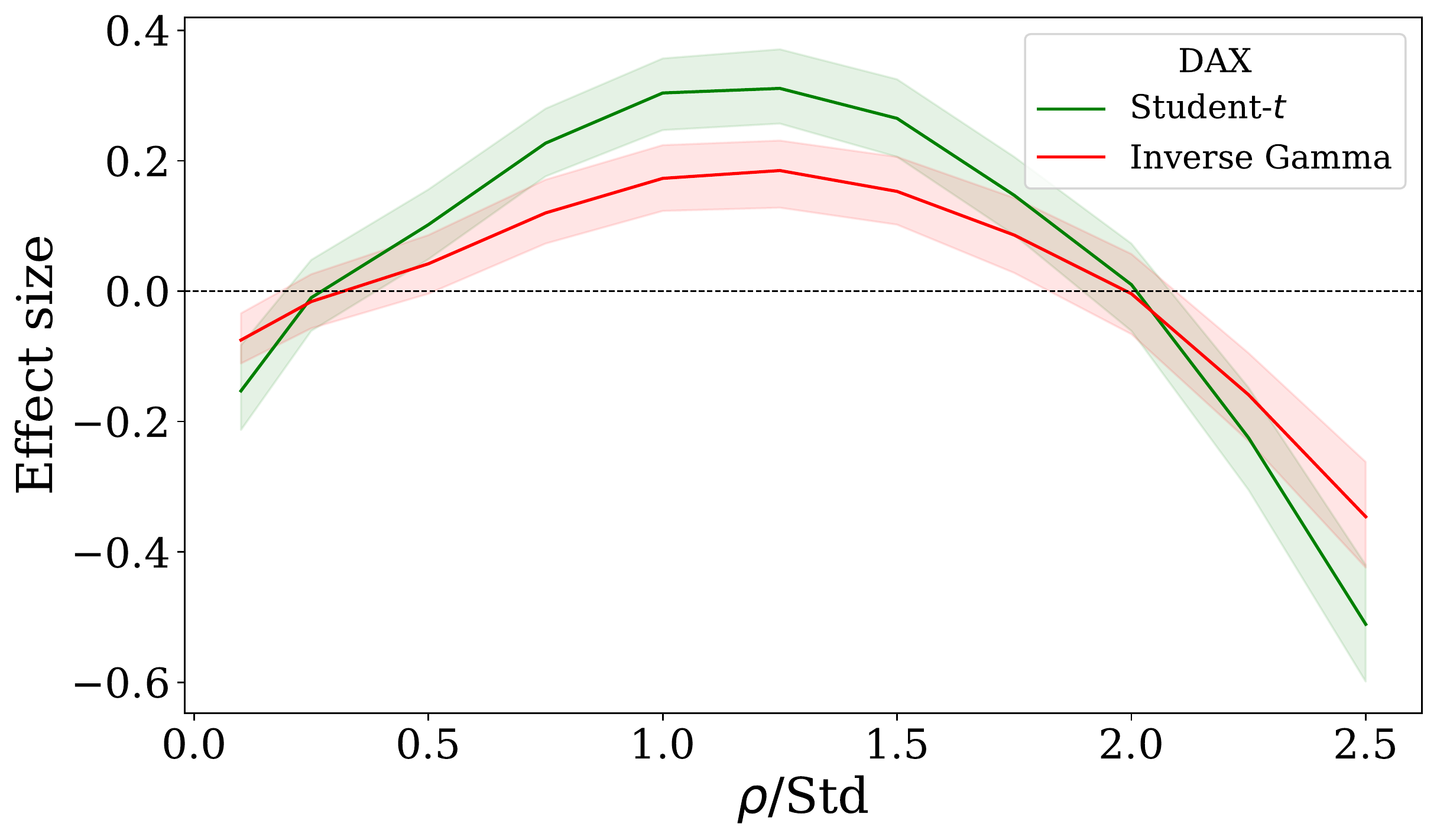}
    \caption{\small{The same as Figure \ref{fig:filter_size_scan} for the posterior distribution of the effect size $d$ as a function of the parameter $\rho/{\rm Std}$. In left panel the Australian index ASX200 while in right panel the German index DAX. }}
    \label{fig:rho_scan}
\end{figure}

\begin{table}[ht!]
    \centering\tiny{
\begin{tabular}{ccccccccccccc}
\toprule
\toprule
$\rho/{\rm Std}$ scale factor [\%]   &    10  &    25  &    50  &    75  &    100 &    125 &    150 &    175 &    200 &    225 &    250 \\
\midrule
ASX200      & -0.126 & -0.103 &  0.087 &  0.185 &  0.259 &  0.331 &  0.371 &  0.383 &  0.352 &  0.233 &  0.024 \\
BVSP        & -0.008 & -0.008 &  0.019 &  0.041 &  0.041 &  0.036 &  0.017 & -0.031 & -0.161 & -0.353 & -0.550 \\
DAX         & -0.153 & -0.010 &  0.102 &  0.227 &  0.304 &  0.311 &  0.265 &  0.147 &  0.010 & -0.225 & -0.511 \\
DJI30       & -0.224 & -0.035 &  0.225 &  0.358 &  0.461 &  0.464 &  0.488 &  0.520 &  0.544 &  0.626 &  0.440 \\
EUROSTOXX50 & -0.211 & -0.020 &  0.167 &  0.315 &  0.373 &  0.449 &  0.525 &  0.475 &  0.450 &  0.283 & -0.059 \\
HSI         & -0.095 &  0.012 &  0.054 &  0.094 &  0.134 &  0.201 &  0.256 &  0.295 &  0.324 &  0.311 &  0.303 \\
NIKKEI225   & -0.126 & -0.053 &  0.071 &  0.090 &  0.064 & -0.010 & -0.104 & -0.253 & -0.426 & -0.590 & -0.756 \\
S\&P500      & -0.268 &  0.004 &  0.279 &  0.422 &  0.477 &  0.470 &  0.499 &  0.603 &  0.617 &  0.531 &  0.279 \\
SHANGAI     & -0.068 & -0.061 &  0.014 &  0.113 &  0.161 &  0.144 &  0.136 &  0.148 &  0.147 &  0.182 &  0.277 \\
VIX         &  0.147 &  0.061 & -0.219 & -0.427 & -0.543 & -0.606 & -0.585 & -0.472 & -0.362 & -0.217 & -0.062 \\
\bottomrule
\bottomrule
\end{tabular}

    \caption{\small{The same as Table \ref{tab:sensitivity_rho_student_t} in the case of the Inverse Gamma hierarchical model.}}
    \label{tab:sensitivity_rho_fht}}
\end{table}
As for the filter size, we omit in both tables the confidence level for the estimation of the mean value. Again, they are roughly included in the range $0.025 \div 0.035$ for all the mean values. We observe a dependency on the parameter $\rho$, in fact all series changes the sign of $d$ over the scan. 
For both the considered hierarchical models we generally observe that for small values of $\rho$ the credible region of $d$ is close to zero, and goes to be negative when $\rho/{\rm Std} \lesssim 0.5$. For larger value of the threshold level $\rho/{\rm Std} \gtrsim 2$, it is hard to find a common pattern. Many indices have a positive credible region, see HSI and SHANGAI, while others are close to zero, such as ASX200 or EUROSTOXX50. For other indices our belief about effect size for large values of $\rho$ is that $d$ is negative: NIKKEI225, BVSP and DAX. The previous statements do not hold in the case of VIX, where the inverse behaviour is observed. Those findings are consistent of those reported in Ref. \cite{jensen2003inverse} where the authors claim that for DJI30 the amount of asymmetry is an increasing function of $\rho$. 
To better illustrate behaviours discussed above we represent in Figure \ref{fig:rho_scan} the results for two interesting cases: ASX200 (left panel) and DAX (right panel). In figure are show the HDI at 94\% of the posterior distribution of $d$ (shadow area) and the posterior mean values (tick lines).

Also in this case we analyze which model is the most probable using the WAIC. In general the Inverse Gamma model is better fitting, however for $\rho/{\rm Std} \leq 0.25$ we observe that the Student-$t$ is favourite for all the indices. As an example for ASX200 and $\rho/{\rm Std} = 0.1$ we have $-10380$ for the Student-$t$ and $-8572$ for Inverse Gamma. This behaviour for small values of the threshold level can be explained by the fact that the long right-tail of the Inverse Gamma distribution cannot fit the shape of $\tau_{\pm}$.

As done for the analysis of the filter size $f$ in Section \ref{fig:filter_size_scan} we check the convergence of the MCMC chains. The Gelman-Rubin statistics are $R_c \simeq 1.01$ for all the chains, thus we can assume that convergence has been achieved. For the ESS parameter we observe a non-trivial dependency on the threshold level. For small values of $\rho/{\rm Std}$ we get ${\rm ESS} \simeq 1000$, while for $\rho/{\rm Std} \simeq 1$ we obtain ${\rm ESS} = 2000 \div 4000$, thus the estimation of the posterior mean of $d$ is less accurate for small values of the threshold parameter. However, it is sufficient for the scope of our analysis.

\subsection{Dynamics effects}
Finally, we study the robustness of our results with respect to the particular time interval considered for the underlying dataset. For similar analysis see for instance Ref. \cite{rodriguez2019multi} where the authors study the dynamic of the asymmetry using trend returns instead of daily returns. This further analysis is important in order to confirm the reliability and the accuracy of our analysis. In particular, for each considered financial time series,  we analyse how the effect size varies by considering a  rolling period of five years starting by 2008, while both the  filter size $f$ and the threshold parameter $\rho$ are fixed respectively to $f=100$ and $\rho=0.028$ (with this value the ratio $\rho/{\rm Std} \in [0.16, 1.61]$).
The mean values of the posterior distribution of the effect size $d$ are reported in Table \ref{tab:sensitivity_time_student_t} for the Student-$t$ hierarchical model and in Table \ref{tab:sensitivity_time_fht} for the Inverse Gamma model.

\begin{table}[h!]
    \centering\small{
    \begin{tabular}{cccccccccc}
\toprule
\toprule
{\bf Year} &   2013 &   2014 &   2015 &   2016 &   2017 &   2018 &   2019 &   2020 \\
\midrule
ASX200      &  0.072 &  0.038 &  0.149 &  0.231 &  0.246 &  0.112 &  0.023 &  0.366 \\
BVSP        & -0.023 &  0.114 &  0.127 &  0.110 &  0.125 & -0.059 & -0.117 & -0.033 \\
DAX         &  0.165 &  0.134 &  0.216 &  0.180 &  0.205 &  0.178 &  0.018 &  0.226 \\
DJI30       &  0.127 &  0.168 & -0.077 &  0.284 &  0.396 &  0.665 &  0.444 &  0.224 \\
EUROSTOXX50 &  0.086 &  0.090 &  0.209 &  0.247 &  0.291 &  0.163 & -0.006 &  0.270 \\
HSI         &  0.023 &  0.048 & -0.041 &  0.095 &  0.085 &  0.274 &  0.171 &  0.104 \\
NIKKEI225   &  0.137 &  0.122 &  0.054 &  0.134 &  0.184 &  0.350 &  0.283 &  0.195 \\
S\&P500      &  0.169 & -0.011 & -0.183 &  0.377 &  0.587 &  0.816 &  0.553 &  0.317 \\
SHANGAI     &  0.132 &  0.113 &  0.076 &  0.069 & -0.017 & -0.076 &  0.022 & -0.028 \\
VIX         &  0.125 &  0.098 &  0.085 &  0.111 &  0.081 &  0.117 &  0.158 &  0.182 \\
\bottomrule
\bottomrule
\end{tabular}}
    \caption{\small Mean values of the posterior distribution of the effect size $d$ defined in Eq. \eqref{eq:effect_size} as a function of the time interval in the case of Student-$t$ hierarchical model. We use a rolling window of five years fixing the filter size $f=100$ business days and the threshold level $\rho=0.028$. }
    \label{tab:sensitivity_time_student_t}
\end{table}

\begin{table}[h!]
    \centering\small{
    \begin{tabular}{cccccccccc}
\toprule
\toprule
{\bf Year} &   2013 &   2014 &   2015 &   2016 &   2017 &   2018 &   2019 &   2020 \\
\midrule
ASX200      &  0.043 &  0.022 &  0.098 &  0.161 &  0.179 &  0.083 &  0.005 &  0.197 \\
BVSP        & -0.017 &  0.088 &  0.100 &  0.060 &  0.066 & -0.057 & -0.085 & -0.027 \\
DAX         &  0.102 &  0.082 &  0.121 &  0.108 &  0.115 &  0.104 &  0.008 &  0.127 \\
DJI30       &  0.083 &  0.106 & -0.055 &  0.182 &  0.260 &  0.383 &  0.232 &  0.091 \\
EUROSTOXX50 &  0.027 &  0.045 &  0.121 &  0.150 &  0.169 &  0.086 & -0.022 &  0.152 \\
HSI         & -0.005 &  0.020 & -0.038 &  0.050 &  0.046 &  0.181 &  0.122 &  0.055 \\
NIKKEI225   &  0.080 &  0.068 &  0.025 &  0.065 &  0.107 &  0.192 &  0.158 &  0.100 \\
S\&P500      &  0.082 & -0.016 & -0.114 &  0.236 &  0.366 &  0.453 &  0.286 &  0.136 \\
SHANGAI     &  0.070 &  0.071 &  0.044 &  0.046 & -0.007 & -0.053 &  0.007 & -0.031 \\
VIX         &  0.052 &  0.047 &  0.016 &  0.053 &  0.019 &  0.020 &  0.059 &  0.077 \\
\bottomrule
\bottomrule
\end{tabular}}
    \caption{\small The same as Table \ref{tab:sensitivity_time_student_t} assuming the Inverse Gamma hierarchical model.}
    \label{tab:sensitivity_time_fht}
\end{table}

\begin{figure}
    \centering
    \includegraphics[width=0.45\textwidth]{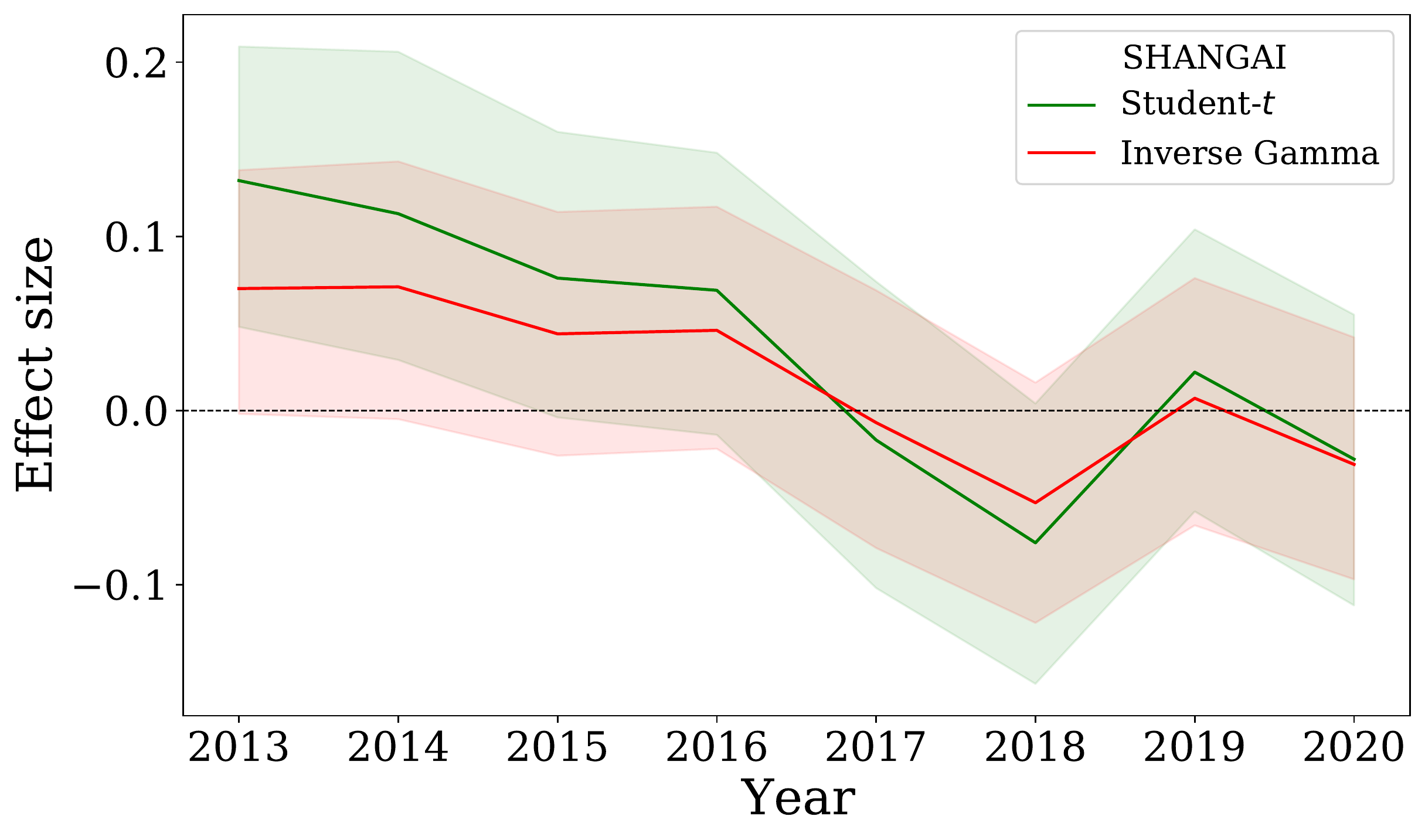}
    \includegraphics[width=0.45\textwidth]{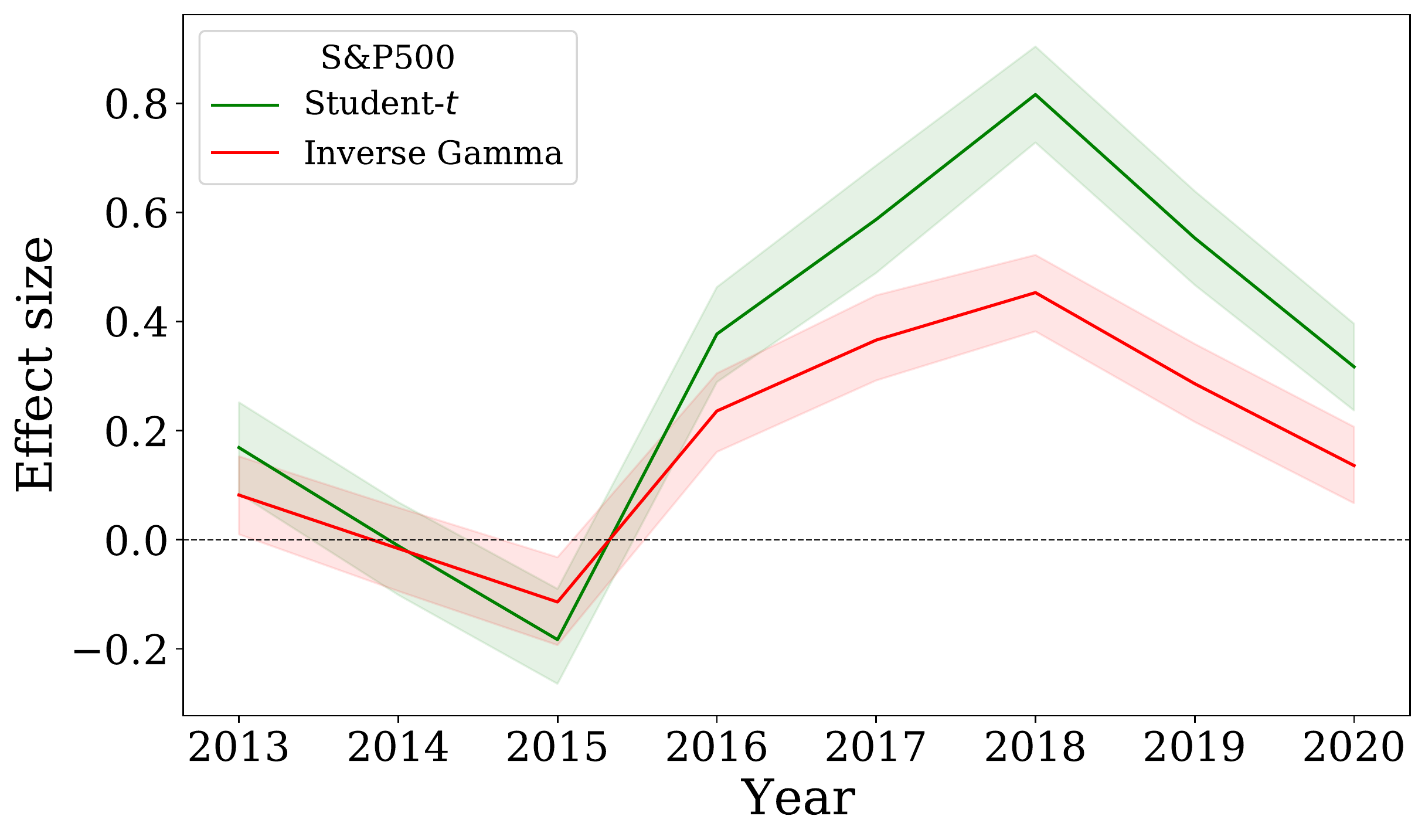}
    \caption{\small The same as Figure \ref{fig:filter_size_scan} for the posterior distribution of the effect size $d$ as a function of the five years rolling window used to estimate the parameters. In the left panel the Chinese index SHANGAI while in the right panel the US index S\&P500. }
    \label{fig:year_scan}
\end{figure}

For the sake of brevity we report only the mean values of the posterior. The error estimate is in the range $0.033 \div 0.080$ for the considered dataset. We note that there exists a strong dependence on the time interval, as expected due to the high non-stationarity of financial time series \cite{cont2001empirical, chakraborti2011econophysics} in which phenomena like volatility clustering that affects different time intervals \cite{ding1996modeling}. In fact, few indices get negative values of the mean even if the whole trend is positive. As an example we show in Figure \ref{fig:year_scan} the evolution of the posterior mean of the effect size $d$ (tick lines) as well as the HDI at 94\% (shadow areas) for two selected indices, the SHANGAI (left panel) and S\&P500 (right panel). Both the hierarchical models are represented, in green the Student-$t$, while in red the Inverse Gamma. It is straightforward to observe that the effect size evolves during time. For instance, the SHANGAI index has a negative trend, while S\&P500 achieve a maximum for (2013-2018) and a minimum in (2010-2015). A non-trivial behaviour is observed in all time-series.

For both the hierarchical models we computed the WAIC, and we get that it is lower in the case of Inverse Gamma model for all the indices except for VIX. In fact, for all the considered time intervals the Student-$t$ model fits better the data. 

We assessed the convergence in the estimation of the effect size $d$ using the Gelman-Rubin statistic $R_c$ discussed in Appendix \ref{sec:appendix_diagnostics}. We get a value close to one for all the configuration analyzed, thus we can safely assume that the estimation on $d$ is well performed. Additionally we compute the effective sample size ESS: for all the considered configuration it is greater in the case of Student-$t$ model and it is constrained in the range $1500 \div 7000$, while in the case of Inverse Gamma $1000 \div 2500$. In any case this is sufficient for our analysis to investigate the behaviour of $d$.

\section{Conclusion}
\label{sec:conclusion}
In this paper we analyzed through a Bayesian framework the widely established gain/loss asymmetry, see Refs. \cite{cont2001empirical, chakraborti2011econophysics, jiang2019multifractal}. Our analysis has been carried out for stock indices with a generalization of the t-Test originally proposed in Ref. \cite{kruschke2013bayesian} in the context of quantitative psychology. We were able to quantify the amount of asymmetry neglecting any normality assumptions on the distribution of $\tau_{\pm}$ using a Student-$t$ hierarchical model. Additionally, we consider a model inspired by the classical results on the first hitting time distribution of the Brownian geometrical motion. In this setting, the likelihood is an Inverse Gamma distribution. The convergence has been achieved for all the MCMCs performed in our analysis, and we generally observe that the Inverse Gamma model better fits the observations thanks to a lower WAIC score. This is a straightforward consequence of the positive definition of the likelihood function. Nevertheless, the Student-$t$ model can also be used to get fast and reliable estimation of the effect size.

We observe that the magnitude of the asymmetry is a non-trivial function, and it depends on several factors: the first one is the considered time-series. For instance, NIKKEI225 and BVSP have a value of $d$ close to zero, while VIX has a large negative value. Moreover, the magnitude is also dependent on the chosen hierarchical model. Generally, the Student-$t$ model is greater (in absolute value) with respect to the Inverse Gamma one. This is related to a small variance $\sigma_{\pm}$ obtained from the posterior distribution.

Other factors that influence the size of $d$ are the time interval as well as two additional parameters: the filter size $f$, used to detrend the time-series, and the threshold level $\rho$ used to define the hitting time. Those parameters have a strong impact on the magnitude, as discussed in Section \ref{sec:section-results}. In particular, we observe that a strong dependence on the considered time interval, this is due to the high non-stationarity of financial time-series. As a consequence correct forecasting of $d$ is difficult to be achieved, see for instance Ref. \cite{takahashi2019modeling} for a recent discussion on the forecasting of financial time-series.

The models presented in this paper are quite simple and can be improved by more complex hierarchical models. In fact, the Student-$t$ can be seen as a benchmark model, while the Inverse Gamma model is a generalization inspired by the stochastic process theory. It is possible to create more feasible models such as inspired by the distribution of the Ornstein-Uhlenbeck hitting time and further research can be done in this direction.


\appendix

\section{MCMC diagnostics}
\label{sec:appendix_diagnostics}
We consider $n_{\rm chains}$ chains of length $n_{\rm draw}$. For a model parameter $\theta$, let $\{\theta_m\}_{m=1}^{n_{\rm chains}}$ be the $m$-th simulated chain. In the following $\hat{\theta}_m$ and $\hat{\sigma}^2_m$ are the sample posterior mean and variance of the chain.



\subsection{Gelman-Rubin statistic}
Let the overall sample posterior mean of the parameter $\theta$ be $\hat{\theta}=(n_{\rm chains})^{-1}\sum_m^{n_{\rm chains}}\hat{\theta}_m$. 
The between-chains $\mathcal{B}$ and within-chain $\mathcal{W}$ variances are defined as
\begin{equation}
    \mathcal{B} = \frac{n_{\rm draw}}{n_{\rm chains}-1}\sum_{m=1}^{n_{\rm chains}}(\hat{\theta} - \hat{\theta}_m)^2 \qquad \mathcal{W} = \frac{1}{{n_{\rm chains}}} \sum_{m=1}^{n_{\rm chains}} \hat{\sigma}^2_m
\end{equation}

It has been shown in \cite{gelman1992inference} that under certain stationary conditions, the pooled variance $\hat{\mathcal{V}}=(n_{\rm draw} -1)\mathcal{W}/n_{\rm draw} + (n_{\rm chains}+1)\mathcal{B}/n_{\rm chains}n_{\rm draw}$ is an unbiased estimator of the marginal posterior variance of $\theta$. The Potential Scale Reduction Factor (PSRF) $R_c$ is proportional to the ratio of $\hat{\mathcal{V}}$ over $\mathcal{W}$ \cite{brooks1998general}. If the $n_{\rm chains}$ chains have converged to the target posterior distribution, then PSRF should be close to 1. PSRF estimates the potential decrease in the between-chains variability $\mathcal{B}$ with respect to the within-chain variability $\mathcal{W}$. If $R_c$ is large, then longer simulation sequences are expected to either decrease $\mathcal{B}$ or increase $\mathcal{W}$ because the simulations have not yet explored the full posterior distribution. In \cite{brooks1998general} it is stated that if $R_c$ is lower than $1.2$ for all model parameters, one can be confident that convergence has been reached. Otherwise, longer chains or other means for improving the convergence may be needed.

\subsection{Effective sample size}
It can be proved under very soft assumptions that MCMC converge to the target distribution as the
number of draws approaches infinity. However, in finite times we need a measure for the estimation error and the goodness of the estimation. Moreover, we recall that in MCMC methods samples will typically be autocorrelated, increasing the uncertainty of the estimation of posterior quantities of interest. The amount by which autocorrelation within the samples increases uncertainty in estimates can be measured by the Effective Sample Size (ESS), which for a single chain of length $n_{\rm draw}$ is defined by \cite{gelman2013bayesian} 
\begin{equation}
 {\rm ESS} = \frac{n_{\rm draw}}{1 + 2 \sum_{t=0}^\infty \rho_{t}}   
\end{equation}
where $\rho_t$ is the autocorrelation at lag $t\geq 0$.
Roughly speaking, the ESS corresponds to  the number of independent samples with the same estimation power as the $n_{\rm draw}$  autocorrelated samples so that the estimation error is proportional to $1/\sqrt{{\rm ESS} }$.

In this paper we use the estimator of the Effective Sample Size (ESS) as defined in \cite{vehtari2020rank},  that is
\begin{equation}
    {\rm ESS} = \frac{n_{\rm chains}n_{\rm draw}}{\hat{\tau}} \qquad \hat{\tau} = -1 + 2 \sum_{t=0}^K \hat{P}_{t}
\end{equation}
where $\hat{P}_t$ is related to the estimated autocorrelation function of the Markov chain $\hat{\rho}_t$ as $\hat{P}_t = \hat{\rho}_{2t} + \hat{\rho}_{2t+1}$. The index $K$ is the last integer for which $\hat{P}_K > 0$.

\bibliographystyle{elsarticle-num}
{\footnotesize
\bibliography{references}
}
\end{document}